# Assessment of Empathy in an Affective VR Environment using EEG Signals

Maryam Alimardani [1*], Annabella Hermans [1] and Angelica M. Tinga [1,2]

[1] Department of Cognitive Science and Artificial Intelligence, Tilburg University, The Netherlands
[2] Department of Human Factors in Vehicle Automation, Institute for Road Safety Research, The Netherlands

\* Correspondence: m.alimardani@uvt.nl

**Abstract:** With the advancements in social robotics and virtual avatars, it becomes increasingly important that these agents adapt their behavior to the mood, feelings and personality of their users. One such aspect of the user is empathy. Whereas many studies measure empathy through offline measures that are collected after empathic stimulation (e.g. post-hoc questionnaires), the current study aimed to measure empathy online, using brain activity collected during the experience. Participants watched an affective 360° video of a child experiencing domestic violence in a virtual reality headset while their EEG signals were recorded. Results showed a significant attenuation of alpha, theta and delta asymmetry in the frontal and central areas of the brain. Moreover, a significant relationship between participants' empathy scores and their frontal alpha asymmetry at baseline was found. These results demonstrate specific brain activity alterations when participants are exposed to an affective virtual reality environment, with the level of empathy as a personality trait being visible in brain activity during a baseline measurement. These findings suggest the potential of EEG measurements for development of passive brain-computer interfaces that assess the user's affective responses in real-time and consequently adapt the behavior of socially intelligent agents for a personalized interaction.

## 1. Introduction

We are currently in an era of rapid technological advancements, which characterizes itself by technologies that merge the physical, psychological and digital spheres [1]. New technologies are rapidly arising in the fields of artificial intelligence, robotics, and virtual reality (VR) impacting all disciplines and industries. With the emerging focus on human-centered innovation, more attention is directed towards socially attentive and adaptive agents [2]. Future robots and virtual avatars are expected to collaborate with humans in different social scenes and to display realistic, human-like and intelligent behavior. In order to facilitate social interaction between humans and agents, it is thus necessary for agents to 'understand' their user. This could be achieved by gaining an insight into the user's affective state during the interaction, computing how the user is feeling, what mood s/he is in and how s/he is responding to an emotional situation. In the present study, the term affective state will be used when referring to emotion and mood-related causes.

In the field of psychology, affective state has been found to have an influence on our memory, assessment, judgement, opinion, expectations and more behavior-related factors [3]. With the use of affective computing it is possible to capture information about the user's affective state, which is considered to be essential for better communication.

Conventional approaches have estimated the affect of the user using offline measures that are collected after the experience, such as post-hoc subjective questionnaires [4]. More recently, affect-related information has been acquired online using facial and vocal recognition and eye-tracking data that are collected during the affective experience [5, 6]. There are also methods that estimate affective states online based on the user's motor behavior, for instance, in the form of mouse and keyboard usage [7]. Moreover, recent trends show an increased interest toward physiological signals and bio-indices of affect captured by such measures as galvanic skin response, heart rate and facial muscle activity [8, 9] and brain activity measurements [10, 11]. The advantage of using physiological measurements is that wearable sensors can record rich and precise data without interruption, with no control of the user. Particularly with brain activity measurements, previous research has shown that real-time systems such as passive brain-computer interfaces (BCIs) can extract mental and emotional states from the user and employ this information to either adapt the interface or provide feedback to the user to improve performance [12, 13]. Similarly, with the passive and continuous neurophysiological input from these BCIs, agents can adapt their behavior based on the human's affective state and enhance engagement during the interaction. Thus, it is important to first identify neuro-correlates of specific affective responses extracted from online recordings as a benchmark for development of passive BCIs.

In addition to the user's affective state, estimating a user's personality traits can also improve human-agent interaction (HAI). One example of a personality trait that could influence an individual's behavior in social interactions is empathy. Empathy is defined as having a similar affective state as another person as a result of the perception of the other person's situation [14]. There are two types of empathy, 1) trait empathy, a person's overall level of empathy, and 2) state empathy, the empathic concern towards a person or stimulus at a particular moment. Through empathy people can share their emotions and create a better understanding about the way of thinking of others. Both trait and state empathy can be divided into two different systems with different neural underpinnings, affective (emotional) empathy and cognitive empathy [15]. For example, one instance of affective (emotional) empathy is the feeling of personal discomfort in response to another person's suffering. Cognitive empathy refers to the ability to understand the cognitive perspective and the state of mind of another individual. In the current study, we focus on trait empathy and affective empathy.

Previous brain imaging studies have investigated the neural basis of empathy in the brain. A large-scale lesion study with 123 participants suffering from a neurodegenerative disease used structural MRI brain volume measurements and found empathic concern to be regulated in the right anterior temporal and medial frontal regions of the brain [16]. Another lesion study demonstrated that patients with brain damage in the prefrontal area of the brain, especially in the right ventromedial cortex, were significantly impaired in empathy compared to patients with posterior lesions and healthy participants [17]. This suggests that empathy is regulated mainly in the frontal and temporal brain regions in the right hemisphere.

Researchers have also conducted studies measuring empathy in the brain with EEG. EEG has an advantage for HAI because the brain activity measurement could be seamless, unobtrusive and be integrated and embedded in devices such as VR headsets. In most of EEG research investigating empathy, a painful stimulus is utilized to evoke empathy towards another person being in pain. For instance, by showing participants images of a hand in a painful situation or in a neutral situation, Mu et al. [18] found increased theta synchronization and increased alpha desynchronization in respectively the frontal/central and the central/parietal brain regions. Even when participants were told that a patient suffering from a neurological deficit would experience pain from a non-painful stimulus and vice versa, power in the alpha band was still found to be suppressed particularly in the frontocentral regions for the non-painful stimuli that were supposedly painful for the patient [19]. A comparable study on brain responses to the pain of others used an immersive CAVE (CAVE Automatic Virtual Environment) environment to display a virtual avatar, which was either sitting on a chair, moving normally or moving and expressing pain. Participants showed greater alpha suppression in the central and temporal brain area for the avatar that was expressing pain versus the avatar that was just moving or was sitting still (neutral). Therefore, alpha suppression could not be explained by movements, but was due to the pain expressions of the avatar [20].

Other studies analyzed EEG power by comparing power in frequency bands over the two cerebral hemispheres of one individual [21]. Most often, this is done by computing the power difference between the left and right hemispheres. An advantage of this method is that it counterbalances individual differences such as skull thickness. In the context of emotion, a large body of literature examined EEG alpha power in the frontal area of the brain, which is generally associated with a state of mind that is awake but relaxed [22, 23]. Since the alpha band is found to be inhibitory, a greater cortical activation is equivalent to a decrease in alpha power [21]. A difference in alpha power between the two hemispheres (asymmetry) in the frontal area was found to be involved in emotion and motivation [24]. Very generally stated, a larger left frontal asymmetry is related to positive affect and a larger right frontal asymmetry is related to negative affect.

Tullett et al. [24] hypothesized that empathy is linked to right frontal EEG asymmetry because empathy evokes a feeling of personal distress as a consequence of sharing the pain or sorrow of other people. They recorded EEG signals while participants were viewing images of charity organizations and collected self-reported sadness, personal distress, perspective taking and empathic concern towards the images. Results showed a positive correlation between right frontal alpha asymmetry and empathic concern, mediated by feelings of sadness. However, in the parietal brain area, a reversed correlation was found with the right parietal asymmetry being negatively correlated to empathy. This could indicate that the right-sided asymmetry is limited to the frontal region of the brain. This also corresponds with the lesion studies discussed before.

The current study was similar to that of Tullett et al. [24] but instead of using still images of charity organizations, an immersive 360° VR video of the same nature was used to induce empathic concern. The reason for this selection of stimuli was two-fold; 1) the VR video was presented in a head-mounted display and therefore any possible visual distraction from the environment was avoided, 2) it was speculated that the immersive VR experience was associated with a higher sense of presence during stimulus presentation and hence would make the empathic responses more pronounced [25]. Moreover, in the previous work, empathy was investigated as an affective state, whereas the current study included empathy as a personality trait, measured by a questionnaire at the baseline.

Two research questions are central to the current paper. First, whether an empathy-inducing VR video influences brain activity measurements, and second, whether there is a correlation between the self-reported level of empathy (personality trait) and the EEG asymmetry in the presence and absence of empathy-inducing stimuli. Based on the research discussed above, we expected to find effects in the frontal and central areas of the brain, specifically in the asymmetry of the alpha band. We also expected empathic participants to have a stronger EEG response to the stimulus versus less empathic participants [26].

## 2. Materials and Methods

### 2.1. Participants

Fifty-two university students (33 female, age $M = 20.3$, $SD = 2.3$) were recruited and received 1.5 course credit for their participation. Participants were included if they were native Dutch speakers and if they reported no current cardiovascular or neurological disorder. The study was approved by the Research Ethics Committee of Tilburg School of Humanities and Digital Sciences.

### 2.2. Materials and apparatus

To measure brain activity, the B-Alert X10 Electroencephalogram was used including nine channels covering the frontal, central and parietal areas of the brain following the 10/20 system of electrode placement (F3, Fz, F4, C3, Cz, C4, P3, POz and P4). In addition, two sensors were placed on the mastoid bones behind both ears as reference electrodes to cancel the noise from the data. The EEG signals were recorded continuously throughout the experiment at 256 samples per second by the software program Acqknowledge 4.4 (BIOPAC Systems Inc.) running on a computer dedicated to collecting the EEG data. Care was taken to keep the scalp-electrode impedance below 40 kΩ for all electrode sites using conductive gel.

The self-reported level of empathy was measured using the Toronto Empathy Questionnaire [27] with a 7-point Likert scale. This questionnaire consists of sixteen empathy-related questions such as "when someone else is feeling excited, I tend to get excited too" or "it upsets me to see someone being treated disrespectfully."

The stimuli were displayed in VR using the HTC-Vive headset (resolution: 1080 x 1200 pixels per eye; refresh rate: 90 Hz, field of view: 110 degrees).

*2.3. Stimuli*

For this experiment, two 360 degrees VR videos were used. The first video was used to make the participant familiar with VR environments removing a VR novelty effect. This video [28] (2.18 minutes) contained several sceneries of the world in a 360 degrees view. The second video [29] (4.43 minutes), which was used to induce empathy, was produced by the charity organization 'Terre des Hommes' and illustrated the life of a Kenyan girl who was being abused as a domestic slave. The video contained five scenes in which the girl did housework as a slave, was scolded and punished for her mistakes, and even got sexually abused by the father of the family.

*2.4. Procedure*

The experiment was conducted in a silent research lab on campus. The participants sat down in a comfortable chair in front of a computer screen and received an explanation about the nature of the experiment. After signing the informed consent form, participants filled out the empathy questionnaire and additional demographic questions using Qualtrics [30]. Next, the EEG sensors were placed on the participants' head and the collection of EEG data began. The VR headset was then placed over the EEG sensor strap and the stimuli were presented. The setup of the EEG device together with the head mounted display is shown in Figure 1.

The experiment consisted of four phases all administered in the VR headset: 1) Familiarization, 2) Pre-baseline (Pre-b), 3) VR experience (VRX) and 4) Post-baseline (Post-b). In the Familiarization phase participants first explored the VR environment with the familiarization video. In the Pre-b phase, participants were asked to look at a fixation cross for 3 minutes while their baseline EEG was being recorded. In the VRX phase, participants watched the empathy-inducing video and EEG signals were obtained. Finally, in the Post-b phase, another 3 minutes of baseline EEG was collected. Participants were instructed to sit as still as possible to avoid noise in the EEG signals but were allowed to move their head and look around the VR environment. Once the experiment was finished, all sensors and the VR headset were removed and participants were debriefed.

*2.5. Data analysis*

The FieldTrip-lite [31] EEG processing toolbox (version 31-08-2018) in MATLAB 9.4 (R2018a) was used to process all EEG data. First, the quality of the EEG data was visually inspected for each individual participant by three researchers using the quality check reports provided by FieldTrip. In this way noisy data files were removed from further analysis.

The remaining EEG recordings were bandpass filtered between 0.5 and 50 Hz to remove all high and low frequency components. Then the data was segmented into the three phases, Pre-b, VRX and Post-b. The Familiarization phase was excluded from the data analysis because it was only meant for training purposes.

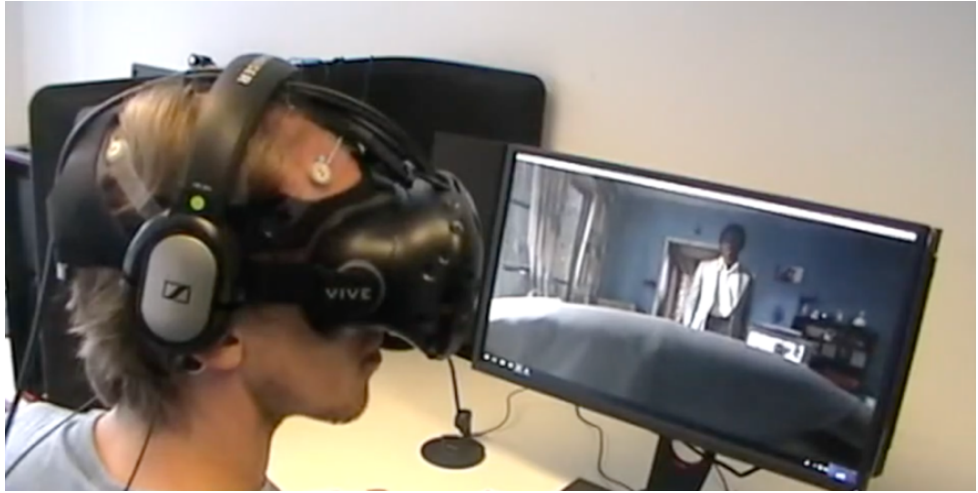

**Figure 1.** Experimental setup; the participant is watching an affective 360° video in a VR headset while EEG signals are recorded.

The remaining EEG recordings were bandpass filtered between 0.5 and 50 Hz to remove all high and low frequency components. Then the data was segmented into the three phases, Pre-b, VRX and Post-b. The Familiarization phase was excluded from the data analysis because it was only meant for training purposes. A Fast Fourier Transform was applied to the EEG segments to compute power spectra and the mean power was obtained for five frequency bands: Delta (0.5 - 3.9), Theta (4 - 7.9), Alpha (8 - 12.9), Beta (13 - 27.9) and Gamma (28 - 50). For every participant, an array of mean EEG powers was constructed for all nine electrodes, in five frequency bands during three phases. The asymmetry powers were then computed by subtracting the left channel from the right channel in frontal, central and parietal areas (i.e. F4 – F3, C4 – C3, and P4 – P3) for each frequency band in each phase.

The data was further analyzed in R [32]. Since the EEG asymmetry data was not normally distributed, non-parametric Kruskall-Wallis tests were used to test if there was a significant main effect between the three phases. This was tested in the three brain areas and for all five frequency bands. Pairwise-comparisons were carried out using Wilcoxon signed-rank tests in those cases where main effects were found.

As for the empathy questionnaire, first participants' answers to the 7-point Likert scale were numerically converted, where never = 0, almost never = 1, rarely = 2, sometimes = 3, often = 4, very often = 5 and always = 6. From the total of 16 questions, half had to be reverse-coded because higher scores indicated less empathy. Next, the sum of scores from all 16 questions was computed and assigned as the total empathy score.

In order to identify the relationship between EEG measurements and subjective responses from the participants, a regression analysis was performed on asymmetry powers and empathy scores. Before performing regression analyses, Cook's distance was used to find the influential observations, where the following formula determined the cutoff threshold for the outliers: 4/N. This was computed for the Pre-b phase to investigate whether a relationship existed before being exposed to the stimulus. It was also computed for the VRX phase to investigate the relationship while the participants were watching the immersive VR video. Moreover, the difference between the Pre-b and the VRX phase was computed to investigate whether participants with a higher empathy score showed stronger brain activity responses to the VR video compared to participants with a lower empathy score.

## 3. Results

After the experimental phase, six participants were excluded due to not correctly following the procedure of the experiment, such as speaking during the video, or technical issues in the experimental procedure. Based on the quality check of the EEG data, another six participants were excluded from the analysis. The data from the remaining 40 participants (14 male, 26 female, age M = 20.53, SD = 2.14) were used for further analysis. Figure 2 visualizes the acquired asymmetry powers over the 5 frequency bands in the frontal (Figure 2a), central (Figure 2b) and parietal (Figure 2c) areas of the brain in each phase of the experiment.

The main effects for the differences between the Pre-b, VRX and Post-b phases were computed for all frequency bands in all brain areas, shown in Table 1. In the frontal brain area, a significant change in the theta and alpha asymmetry was found over the three phases. In the central brain area a significant change in delta and theta asymmetry was found, whereas in the parietal area of the brain, no significant changes were found in any of the frequency bands. Effect-size was computed using the Phi coefficient ($\varphi$), which is the square root of Chi-Squared ($\chi^2$) divided by the number of observations ($\sqrt{(\chi^2/n)}$). The effect-size is considered large when it is .5 or higher, which was solely the case for frontal alpha asymmetry power ($\varphi$ = .66). Since in all measurements, the number of observations n was equivalent, the value of Chi-Squared in Table 1 is representative for comparison of effect-sizes between different locations and frequency bands.

To investigate in which specific phases (Pre-b, VRX and Post-b) effects occurred, a post-hoc test was conducted on the alpha and theta band in the frontal area and for the delta and theta band in the central area (Table 2)

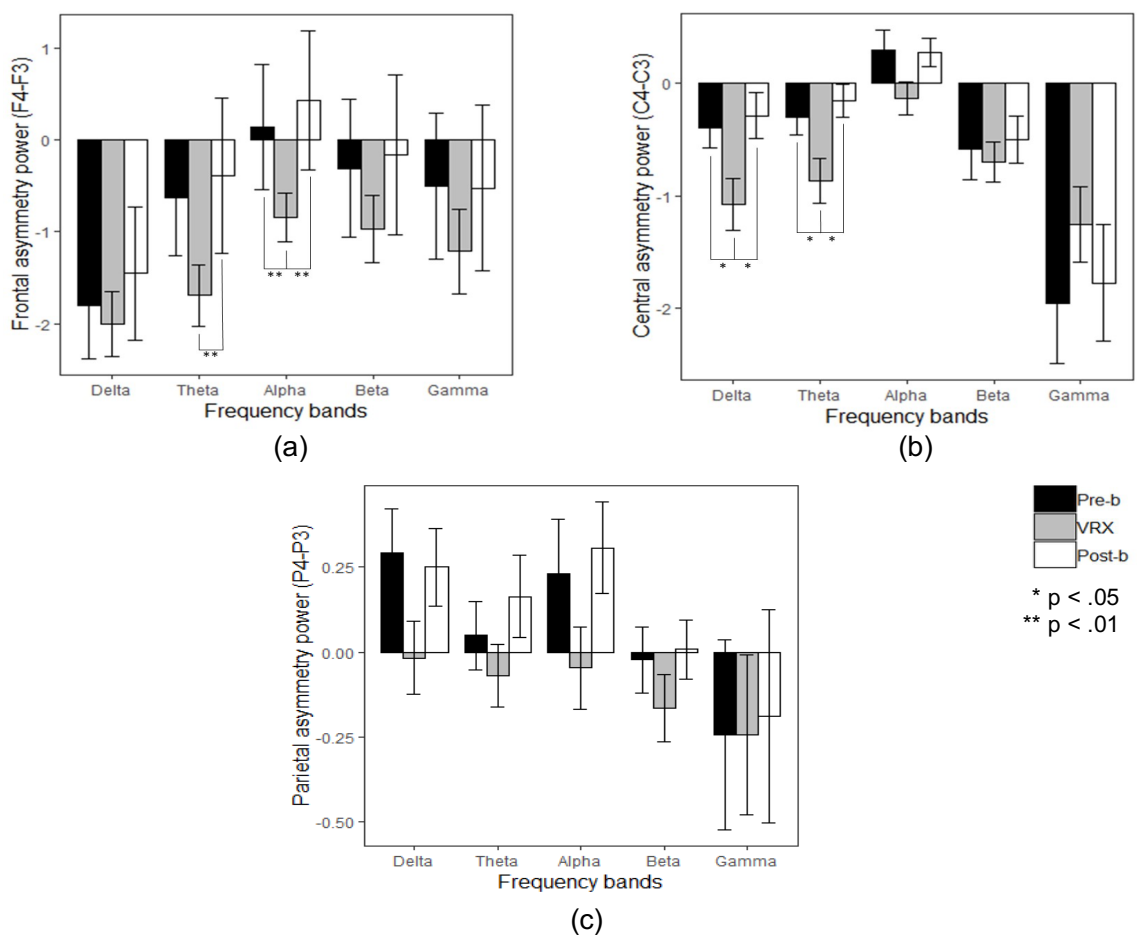

**Figure 2.** Average asymmetry values for five frequency bands in (**a**) frontal, (**b**) central, and (**c**) parietal areas. Error bars represent the standard error of the mean.

Table 1. Results of the Kruskall-Wallis tests testing the difference between the three experimental phases for each brain area and frequency band (significant *p*-values marked in bold).

| Frequency band | Frontal area | | Central area | | Parietal area | |
|---|---|---|---|---|---|---|
| | $X^2(2)$ | $p$ | $X^2(2)$ | $p$ | $X^2(2)$ | $p$ |
| Delta | 0.542 | .762 | 8.320 | **.016** | 4.644 | .098 |
| Theta | 9.512 | **.009** | 9.348 | **.009** | 2.625 | .269 |
| Alpha | 15.5 | **<.001** | 5.948 | .051 | 3.363 | .186 |
| Beta | 2.2 | .54 | 0.805 | .669 | 1.715 | .424 |
| Gamma | 0.553 | .758 | 1.136 | .567 | 0.285 | .867 |

Table 2. Pairwise comparisons between the three experimental phases using Wilcoxon rank sum test (significant *p*-values marked in bold).

| Brain area | Freq. band | Pre-b vs. VRX (*p*) | VRX vs. Post-b (*p*) | Pre-b vs. Post-b (*p*) |
|---|---|---|---|---|
| Frontal | Theta | .068 | **.009** | .214 |
| | Alpha | **.001** | **<.001** | .407 |
| Central | Delta | **.049** | **.021** | .413 |
| | Theta | **.034** | **.011** | .470 |

In the frontal (theta and alpha bands) and central (delta and theta bands) brain areas, significant changes were found from the Pre-b to the VRX phase and from the VRX phase to the Post-b. The only exception was in the theta asymmetry in the frontal area of the brain where a significant change was only found from the VRX phase to the Post-b.

Regression analyses were next conducted only for the frontal alpha asymmetry powers as this was the only significant effect with a large effect size. For the Pre-b and the VRX phases, the frontal alpha asymmetry powers were used in linear regression paths with the empathy scores. Also, the difference between the VRX and the Pre-b (VRX – Pre-b) was computed and used in a linear regression with the empathy scores. After Cook's distance detection, respectively 38, 39 and 37 participants remained for the analysis of the Pre-b, VRX and VRX – Pre-b phases. The statistical results of the linear regression are shown in Table 3. Only in the Pre-b phase, a significant relationship was found between empathy and frontal alpha asymmetry, as shown in Figure 3.

Table 3. Results of the regression analyses testing the relationship between the frontal alpha asymmetry in each experimental phase and empathy (significant findings marked in bold).

| Phases | b | β | T(df) | p |
|---|---|---|---|---|
| Pre-b | 67.507 | -5.687 | -2.619(36) | **.013** |
| VRX | 66.954 | -0.401 | -0.242(37) | .810 |
| VRX – Pre-b | 68.888 | 2.502 | 1.152(35) | .257 |

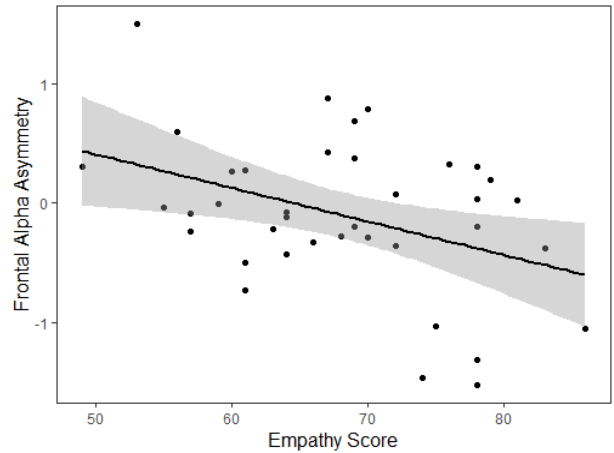

**Figure 3.** Linear regression model of the empathy score and the frontal alpha asymmetry, *p* < .05.

## 4. Discussion

The aim of this study was to assess affective empathy in participants through EEG as online physiological measure in order to contribute to the enhancement of human-agent interaction. Empathy was elicited by showing participants an affective VR video about child exploitation. Before, during and after showing the VR video, participants' brain activity was measured using EEG sensors.

Findings showed that overall EEG asymmetry, defined as the difference in EEG spectral band power between right and left hemispheres, decreased when participants were experiencing the VR video, and increased after the video. This indicates that on average brain activity shifts due to the stimulus, in almost all frequency bands and in all brain areas. Significant transitions in the theta and alpha asymmetry were found in the frontal brain area. Also, in the central area, significant asymmetry changes were found in the lower frequency bands; delta and theta. The parietal area did not show any significant asymmetry transitions.

In the frontal brain region, the alpha asymmetry power shifted from a positive value to a negative value indicating a change of activity over hemispheres during the VR experience. The alpha wave is inhibitory, which means that a negative alpha asymmetry during the VR phase was caused by a greater cortical activation in the right hemisphere. This is consistent with previous literature that indicated that empathy is found to be in the right frontal brain area [15, 16, 23].

Further analyses showed a significant correlation between the frontal alpha asymmetry and empathy during the baseline phase. The frontal alpha asymmetry ranged from negative values for higher empathy scores to positive values for lower empathy scores. This demonstrates that participants with a higher score on empathy scale showed relatively larger cortical activity in the right hemisphere and participants with a lower empathy score had a relatively larger cortical activity in the left hemisphere during the Pre-b measurement. This result indicates that there is a cortical difference between people with different personality characteristics with regards to empathy, even when they are not exposed to the stimulus yet. Similar results were shown in previous research of Tullett et al. [24] where larger right frontal activity was found to be a predictor for empathic concern. This also supports lesion studies discussed earlier, that claim empathy as a trait is related to right frontal regions [15, 16].

We expected that participants with a relatively higher empathy score would have been more influenced by the empathy-inducing immersive video and hence showed a larger transition in their brain asymmetry from the Pre-b phase to the VRX phase compared to the participants with a relatively lower empathy score. However, our results did not support this hypothesis. Even though the asymmetry powers changed significantly in some frequency bands in the frontal and central areas of the brain, the regression analysis did not find a significant relationship between this change and participants' empathy scores. One possible explanation could be that participants did not respond to the immersive video in the same way and hence the changes in the their brain activity differed based on the level of presence they felt in the environment [33]. Future research should look at the correlations between sense of presence in VR and changes in asymmetry powers that participants exhibited in different brain regions. Another possibility is that since the empathy scores were normally distributed in a limited range (obtained score range 49/86, min/max score 0/96), there was not enough variety within the participants to find a relationship with the brain asymmetry alterations. This could be investigated in future research by examining participants that exhibit a very low level of empathy, which could be due to cortical impairments, versus participants that score extremely high on empathy questionnaire.

A possible limitation of the employed analysis in this study is the effect of ocular artifacts in the EEG signals. Although it is a common practice in neuroscientific research to correct eye blinks artifacts in EEG signals through regression-based techniques or independent component analysis (ICA) [34], we did not perform such artifact rejection in our pre-processing. The reason for this was two-fold; 1) we were interested in identifying EEG features that could be detected by real-time BCI systems in the future. Automatic detection and filtering of eye artifacts during recording is not computationally efficient for BCIs [35] and therefore most passive BCI algorithms rely on temporally observable changes in unprocessed EEG signals. In this case, we collected pre- and post-baselines with open eyes and we compared those segments with the VR viewing condition in which the eyes were open, 2) based on the findings of the past research, we were particularly interested in brain activity patterns in the frontal area of the brain. Eye artifact removal, either performed manually as in ICA or by automatic-selection algorithms, can largely affect frontal EEG data and lead to information loss. In order to avoid such possible information loss at the frontal sites, we decided to limit the pre-processing to bandpass filtering of high and low frequency components.

Another limitation of this study is generalizability of our findings to different age groups. Previous studies have shown that age has a significant impact on EEG measures such as spectral power and coherence [36-38]. In this experiment, we recruited our participants from a homogenous age group (around 20 years old) in order to minimize the effect of individual differences on the EEG features. Future research should replicate this study with other age groups including middle-aged and elderly to investigate the general conformity of our results.

An important question regarding our findings is that whether the lateralization of cortical activity during the VR experience is a signature of empathic response and not simply an arousal response to stress. In this study, we focused on affective (emotional) empathy, which is the emotional response towards another person's suffering (here an African child being exploited) and inherently involves personal distress and sadness as a result of empathic concern [15, 24]. Images of charity campaigns have been used in past studies [24] and similar results have been reported for frontal asymmetry as an objective indicator of emotions and approach-withdrawal responses [39-41]. Overall, literature shows that positive emotions such as happiness, pleasantness and interest that are representative of approach or motivational effect are accompanied by increased left frontal activity in alpha band, whereas fear, disgust, anger and sadness, which are associated with withdrawal, lead to increased right frontal activity [24, 39]. Our study provides consistent findings where empathic concern during the VR viewing led to a withdrawal effect and subsequent shift of activation in alpha band from the left to right hemisphere. Future research should further evaluate whether this empathic emotional response can also promote a change of behavior (e.g. taking action to help).

Lastly, future research should examine the effect of different media usage and degrees of immersion to induce empathy. In this research, we used an immersive 360° video together with a VR headset in order to dissociate the participants from the experimental room and induce a sense of presence as if they were in the same location of the African girl [33]. Past research has shown that a greater immersion results into a higher sense of presence, which is in correlation with stronger emotional experience in an affective VR environment [42]. The active viewing of the VR environment in our experiments, in which participants could move their head and watch different corners of the room, was a novel aspect compared to the experimental setup of previous studies where participants only passively watched a video or a series of images on a computer screen (e.g. [15], [24]). Yet, an even higher sense of presence could be achieved by using a VR environment in which users are provided with the ability to interact with the environment and manipulate objects. It is, however, noteworthy that 360° videos are the most widely used and cost-effective way of presenting VR to people and have been used in past research as visual stimuli as well [43].

What do our findings mean for the development of future interactive technologies? With developments in artificial intelligence, social robotics, VR and serious gaming, the use of socially intelligent agents that take into account user responses becomes increasingly important. Whereas agents such as Microsoft's paperclip 'Clippy' [44] did not care too much about the wellbeing of its user, agents such as the intelligent tutoring system AutoTutor [45] took into account individual differences in the verbal input. An affective version of AutoTutor considered the cognitive state of the user by adjusting the pedagogical state of its expressions on the user's affect [46]. Today's agents can take the interaction with the user even a step further by considering empathy. With the availability of sensing technologies and AI algorithms that measure and predict physiological and cognitive responses of the user [38], the next generation of agents is likely to become more interactive and show more empathy with the user.

Another potential application of our findings is in the development of neurofeedback approaches and brain-computer interfaces that will help patients who have limited empathic understanding due to neurological disorders. This has been tested in the past in form of an interactive narrative paradigm where subjects supported a virtual character by means of their brain activity [47] or in forensic psychiatry, where BCI-based empathy training in VR is suggested for the improvement of criminals' emotional responses and suppression of their violent behavior [48].

We expect that applying EEG in combination with interactive technology will become more feasible and common in the future due to technological advancements. For example, EEG sensors will become more comfortable for the user and less costly. Wireless and portable EEG systems are already available which offer increased mobility during data collection. Also, software tools are being developed that aid in processing and classifying EEG signals. Our work shows that such commercial EEG systems together with software tools for data analysis provide sufficient resources to obtain neural indicators of affective experiences in VR. This can set an example for other researchers in the field of human-technology interaction to similarly employ these applications for evaluation and optimization of their interaction protocol. Future VR systems could also include EEG sensors embedded in the headset and have the computational capacity to extract a user's neural changes in real-time and adapt the VR environment to the user's needs. The current study has taken a step toward identifying such EEG features that represent empathic state changes and empathy as a personality trait.

In future studies, other personality traits than empathy could be investigated in relation to particular brain activity patterns. Knowledge about these traits in relation to EEG measurements could help human-agent interaction become more efficient and more adaptive, enhancing the communication and behavior of the agent towards the user.

## 5. Conclusions

This study evaluated the neurophysiological underpinnings of an empathic experience induced by an affective VR environment. We collected EEG signals before, during and after participants watched a video of child exploitation in a head-mounted display. Our results showed significant decreases in alpha, theta and delta asymmetry in the frontal and central brain regions during the affective VR experience. Furthermore, a significant relationship was found between empathy as a personality trait and frontal alpha asymmetry at the pre-baseline indicting that participants with higher empathy scores generally exhibited more cortical activation in the right hemisphere. These findings are useful in the field of wearable sensors and human-technology interaction, where a user's traits and real-time affective responses can be detected by artificial agents and hence improve their interactive behavior.

**Author Contributions:** conceptualization, M.A.; methodology and experiment design, all authors; data acquisition and analysis, A.H.; writing—original draft preparation, A.H.; writing—review and editing, all authors.

**Funding:** This research was partially funded by the European Union, OP Zuid, the Ministry of Economic Affairs, the Province of Noord-Brabant and the municipalities of Tilburg and Gilze Rijen (PROJ-00076). The usual exculpations apply.

**Acknowledgments:** We would like to thank Dylan Tjon who helped with the data collection.

**Conflicts of Interest:** The authors declare no conflict of interest.